\documentclass[aps,showpacs,prl,twocolumn]{revtex4}
\usepackage{amsmath,amssymb,graphicx,bm}
\begin{document}



\title{Metallic Carbon Nanotubes as High-Current-Gain Transistors}

\author{
F. Green${}^{1}$ and D. Neilson${}^{1,2,3}$\\
${}^1$ {\em School of Physics, The University of New South Wales,
Sydney, New South Wales 2052, Australia.}\\
${}^2$  {\em Dipartimento di Fisica, Universit\`a di Camerino,
I-62032 Camerino (MC), Italy.}\\
${}^3$  {\em NEST CNR-INFM, I-56126 Pisa, Italy, Italy.}\\
}

\begin{abstract}
Low-dimensional metallic transport is at
the heart of modern high-performance transistors. While the best devices
are currently based on III-V heterojunction quantum-well channels,
we demonstrate that much greater performance gains reside
in the unique properties of one-dimensional
carbon nanotubes.
Using recent experimental data, we show how the specific
features of degenerate carrier kinetics in
metallic nanotubes enable a novel class of quantum-confined
signal detector whose current gain exceeds present
transistor technology. We analyze the key interplay of degeneracy
and scattering dynamics on the transconductance via
a microscopically conserving quantum-kinetic description.
\end{abstract}




\maketitle



In this paper we explore metallic CNTs as a novel class
of analog current amplifiers, which are core components
of nanoscale electronic design
\cite{review,burke}.
Applying quantum kinetics to the observed current response
of single-wall nanotubes,
we map the current as a function not simply of the driving
voltage along the tube but also, crucially,
as a function of carrier density.

The touchstones of performance in
high-current-gain transistors have been devices
taking advantage of the properties of the quantum-confined metallic
electron gas formed at epitaxial III-V heterojunctions
\cite{HEMT}.
These two-dimensional quantum devices dominate millimeter-wave
signal processing technology thanks to three superior features:
(a) the high carrier density attainable in a quantum-well confined
metallic channel relative to conventional bulk doping;
(b) the ability to sensitively tune the density by simple gate control; and
(c) reliable low-resistance contacts at the channel's source and drain.

Carbon nanotubes hold promise for a novel class of 
nanoscopic amplifier,
reliant upon the unique behavior of one-dimensionally confined
carriers. In broad concept, metallic CNTs share much with their
heterojunction predecessors. Physically, however,
they differ markedly in
much stronger confinement and larger energy scale,
far smaller dimensions, and unmatched stability
\cite{infineon}.

Transistor operation in semiconducting CNT structures is
already well demonstrated
\cite{review,ibm_dns,infineon}.
While both metallic and semiconducting nanotubes share, as
fundamental attributes, quantum confinement and
mechanical and thermal toughness, they differ greatly with
respect to the criteria (b) and (c) above.

One reason why semiconducting CNT devices
do not perform to their theoretical potential is the
difficulty of fabricating source and drain contacts
that are both reproducible and have intrinsically
low access resistance. This degrades the maximum achievable
current-voltage response as well as the signal gain
\cite{review}.
In metallic CNTs, it is known that
these drawbacks can be largely resolved.
This invites a sharper theoretical analysis
of their possibilities.

As far as analog signal detection and amplification are
concerned, developments in metallic CNT transport offer
solid evidence that metallic nanotubes can
outperform their semiconducting counterparts. In so doing, they
should certainly also surpass the best established
III-V quantum-well approaches.
Further, III-Vs and similarly complex technologies cannot be
integrated into true sub-mesoscopic structures; but
the CNTs' rugged simplicity overcomes this obstacle
\cite{review}.

After many experimental studies on transport in CNTs,
it is now practical to fabricate
reproducibly good ohmic contacts to the channel.
These developments have yielded new and
reliable data on electrical response in metallic samples
\cite{pop,pop0,pop1}.
At the same time, kinetic models for CNTs
have begun to probe phonon scattering
effects at voltages well above the linear regime
\cite{ibm_dns,ibm,yao}.

To assess metallic CNTs as super-sensitive analog transistors,
we must predict their current gain. We do this through a
detailed kinetic approach. The description (i) is easily
integrated with current-voltage measurements
\cite{pop}, (ii) allows straightforward extraction of basic scattering
parameters, and (iii) makes no {\em ad hoc} assumptions foreign
to the known behavior of the electron gas.

Consider the electron population in the lowest one-dimensional
conduction subband of a uniform, metallic single-wall nanotube.
Its steady-state momentum distribution $f_k(\mu)$
depends on chemical potential $\mu$ and bath
thermal energy $k_{\rm B}T$. The quantum transport equation,
at external driving voltage $V$ over length $L$, is
\cite{gtd}

\begin{widetext}
\vspace{-5mm}

\begin{eqnarray}
{eV\over \hbar L} {{\partial f_k}\over {\partial k}}(\mu)
&=& -{1\over {\tau_{\rm in}(E_k)}}
{\left( f_k(\mu) -
{{\langle \tau_{\rm in}^{-1} f(\mu) \rangle}\over 
 {\langle \tau_{\rm in}^{-1} f^{\rm eq}(\mu, k_{\rm B}T) \rangle}}
f^{\rm eq}_k(\mu, k_{\rm B}T)
\right)}
- {1\over {\tau_{\rm el}(E_k)}}
{ {f_k(\mu) - f_{-k}(\mu)}\over 2 }.
\label{eq1}
\end{eqnarray}
\end{widetext}

\noindent
The right-hand-side collision terms encode the scattering dynamics
through two relaxation rates: $1/\tau_{\rm in}$
(inelastic) and $1/\tau_{\rm el}$ (elastic). In general, these
vary with subband energy $E_k$.
Traces ${\langle \cdot \cdot \cdot \rangle}$ represent expectations
over wavevector space, spin, and a twofold ``valley'' degeneracy
\cite{review}.
Thus, the expectation of $f_k/\tau_{\rm in}(E_k)$ is
${\langle \tau_{\rm in}^{-1} f \rangle} \equiv
4\int (dk/2\pi) \tau_{\rm in}(E_k)^{-1} f_k$.
The equilibrium function $f^{\rm eq}(\mu, k_{\rm B}T)$ has the usual
Fermi-Dirac form, explicitly dependent on $\mu$
and $k_{\rm B}T$.

Equation (\ref{eq1}) automatically respects
detailed balance. It is unconditionally gauge
invariant. This is due to the explicit form of the collision
terms.
For collision times independent of $E_k$,
Eq.\ (\ref{eq1}) has an analytic solution 
at both low and high fields
\cite{gtd,shsw}.

The core channel is taken to be uniform and defect-free. There
is no internal elastic scattering; it occurs at the interfaces.
Thus  
the elastic mean free path (MFP)
spans the active length of the structure:
$\lambda_{\rm el} = L$.

The characteristic velocity of the system at density
$n = \langle f^{\rm eq}(\mu, k_{\rm B}T) \rangle$ is
its group velocity
at the characteristic wave vector
$k_{\rm av} \equiv
\langle 2|k| f^{\rm eq}(\mu, k_{\rm B}T) \rangle/n$.
This yields
$v_{\rm av}(\mu)\!=\!\hbar^{-1}[dE/dk]_{k=k_{\rm av}}$.
Note that $k_{\rm av}$ and $v_{\rm av}$ go to their
Fermi values $k_{\rm F}$ and $v_{\rm F}$ at large $n$.
The elastic collision time 
is then fixed by
$\tau_{\rm el} = \lambda_{\rm el}/v_{\rm av}(\mu) = L/ v_{\rm av}(\mu)$.
Elastic scattering is insensitive to field effects.

If inelastic collisions act solely at the leads,
transport is ballistic. However, electrons energized
by the field emit longitudinal optical phonons copiously.
The inelastic MFP, $\lambda_{\rm in}(\mu, eV)$,
is not just shorter than $L$ at room temperature (even at low fields)
but depends sensitively on how the carriers' energy gained in transit
is dissipated: the more the carriers draw from the voltage,
the more they shed by enhanced emission
\cite{pop,ibm}.

To extract the physics of
$\lambda_{\rm in}(\mu, eV)$ we start with the solution for
the current, following from Eq.\ (\ref{eq1}).
A metallic single-wall nanotube (MSWNT) has linearly dispersive bands.
For electrons,
$E_k = \hbar v_b |k|$ with $v_b = 5\times 10^7{\rm cms}^{-1}$
\cite{ibm};
for holes, $E_k = -\hbar v_b |k|$.
The expectation $I = \langle -ev f \rangle$ is best evaluated
in the Fourier space dual to $k$. Some algebra yields
\vspace{-3mm}

\begin{eqnarray}
I
&=& 4G_0 V
{\tau k_{\rm F}\over \tau_{\rm el} k_{\rm av}}
{\left[ 1 -
\exp{\left( -{\hbar L k_{\rm av}\over eV\sqrt{\tau\tau_{\rm in}}} \right)}
\right]}
\label{eq4}
\end{eqnarray}

\noindent
where $G_0 \!=\! e^2/\pi\hbar$.
The total relaxation time is
$\tau^{-1} \!=\!\tau_{\rm el}^{-1} \!+\! \tau_{\rm in}^{-1}$.
The expression for $I$ is exact in the degenerate regime,
and remains accurate at low density.

It is interesting to look at the behavior of Eq.\ (\ref{eq4})
at high fields. Suppose $\tau_{\rm in} \sim V^{-\beta}$.
At large $V$, inelasticity dominates the Matthiessen rate:
$\tau^{-1} \to \tau_{\rm in}^{-1}$. When $eV \gg E_{\rm av}$,
there are two possibilities. If $\beta \leq 1$ then
for $A$ a constant 
\vspace{-3mm}

\[
I = nev_b\sqrt{\tau\over \tau_{\rm in}}
{{1 - \exp(-AV^{\beta-1})}\over AV^{\beta-1}}
\to nev_b\sqrt{\tau\over \tau_{\rm in}}
\] 

\noindent
and the current saturates.
Otherwise, for $\beta > 1$ we get
\vspace{-3mm}

\[
I \sim V^{1-\beta} \to 0.
\]

\noindent
The current, which must vanish linearly as $V \to 0$, will
also die off at large $V$. There is a critical field at which $I$
peaks, and then falls again.

\begin{figure}
\vspace{5mm}
  \centering
  \includegraphics[height=60mm]{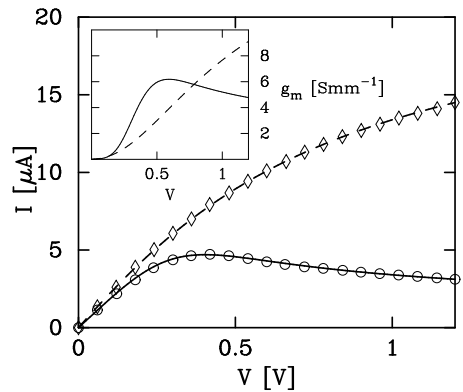}
\vspace{-17mm}
  \caption{\label{f1}
Current $I$ and predicted transconductance $g_m$ (inset)
as functions of source-drain voltage in two types of metallic
single-wall carbon nanotubes.
Channels are identical; only their heat-sinking differs.
Data are from Fig.\ 2(a) of Pop {\em et al.}
\cite{pop}:
{\em diamonds} for the heat-sunk tube,
{\em circles} for the suspended tube.
Curves are our kinetic simulations: {\em dashed} lines for
the heat-sunk and {\em full} lines for suspended tubes.
The radical changes in $I$--$V$ are faithfully reproduced.
}
\vspace{-10mm}
\end{figure}

Exactly that feature marks the response of a suspended MSWNT.
Figure 1 exemplifies this remarkable behavioral change in an actual device.
Two current-voltage ($I$--$V$) characteristics
measured by Pop {\em et al.}
\cite{pop},
are contrasted and compared with our corresponding
calculations: one for a substrate-backed
(and therefore heat-sunk) tube, and the other for
an identical but free-standing tube. Our predicted transconductances
$g_m$ (the current gain) are also shown;
at a driving field of 4kVcm${}^{-1}$ the heat-sunk
nanotube attains a $g_m$ of $9{\rm Smm}^{-1}$ per unit channel width.

Despite poor heat dissipation, the
suspended device promises peak values of
$6{\rm Smm}^{-1}$.
This easily outdoes
the most advanced heterojunction transistors based
on two-dimensional metallic InGaAs quantum well channels,
with a peak $g_m$ of 1.6Smm${}^{-1}$
\cite{HEMT}.
Clearly, CNTs more than hold their own
relative to III-V technology.

The key to the striking qualitative difference observed in the
behavior of $I$ is the $V$-dependence of the inelastic time
$\tau_{\rm in}(\mu,V) = \lambda_{\rm in}(\mu,V)/v_{\rm av}(\mu)$.
Since it so sharply fixes the functional form
of the current, we must look at the asymptotic behavior
in the measured data
\cite{pop}.

\begin{figure}
  \centering
  \includegraphics[height=60mm]{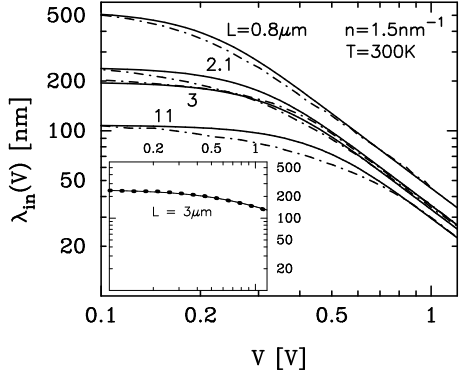}

\vspace{-13mm}
  \caption{\label{f2}
Functional behavior of the inelastic mean free
path in suspended metallic nanotubes.
Chain lines: $\lambda_{\rm in}$ extracted from
data
\cite{pop}.
Full lines: simulation (see Eq.\ (\ref{eq3}) in text).
All curves show a clear $V^{-1.5}$ falloff at high field.
The inset shows $\lambda_{\rm in}$ for
the heat-sunk sample of Fig.\ 1: full dots are points
extracted from data \cite{pop}; full line is our simulation.
}
\vspace{-5mm}
\end{figure}

In Fig.\ 2 we plot the $V$-dependence of the inelastic MFPs,
extracted directly from published measurements; see Fig.\ 3(a) of
Pop {\em et al.}
\cite{pop}.
We do this by inverting Eq.\ (\ref{eq4}) and solving for
$\lambda_{\rm in}(\mu,V)$, given the $I$--$V$ data. Since no value of
carrier density was published with the data, we assume
$n = 1.5{\rm nm}^{-1}$, equivalent to $E_{\rm F} = 400$meV.
In the degenerate regime, the outcome is not sensitive to $n$.

For the heat-sunk device of Fig.\ 1,
the high-field data obey a power-law falloff with exponent
$\beta = 0.75$. For all the suspended samples studied in Ref.
\onlinecite{pop},
there is a very clear falloff with $\beta = 1.5$.
In the low-field regime, $\lambda_{\rm in}$ for each sample
goes to a constant.

With the extracted results we can reliably model
the strong field-dependent inelastic effects
by the analytic fit

\vspace{-4mm}
\begin{equation}
\lambda_{\rm in}(\mu, eV) 
= \lambda_{\rm in}^{(0)}
{{\tanh (u^{\beta})}/u^{\beta}};
{~~~}u = \alpha{eV/ E_{\rm av}},
\label{eq3}
\end{equation}

\begin{figure}
\vspace{2mm}
  \centering
  \includegraphics[height=56mm]{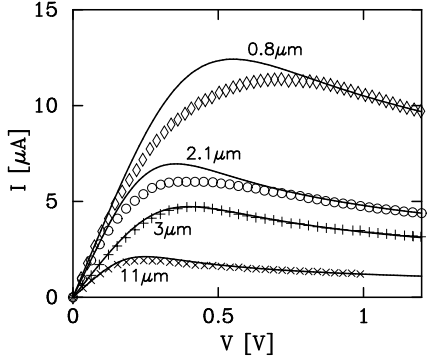}
\vspace{-11mm}
  \caption{\label{f3}
Current response in suspended MSWNTs of 
varying lengths; see also Fig.\ 2.
Symbols are measured data points [Fig.\ 3(a), Ref.
\onlinecite{pop}].
Full lines are our simulation.
}
\vspace{-6mm}
\end{figure}

\noindent
from which $\tau_{\rm in} = \lambda_{\rm in}(\mu, eV)/v_{\rm av}(\mu)$. 
The low-field MFP $\lambda_{\rm in}^{(0)}$ is
set by the low-field slope of the current-voltage curve;
parameters $\alpha$, $\beta$ are set by
$I(V)$ measured at large $V$.
The characteristic energy is $E_{\rm av} \equiv E_{k_{\rm av}}$.

There is a close match between our model $\lambda_{\rm in}$
and its empirical behavior.
At high field, the asymptotic tail of Eq.\ (\ref{eq3})
correctly reproduces the experimental $I$--$V$ results
(see Fig.\ 3). At low field, it goes to the appropriate
inelastic MFP determining the linear slope at $V \to 0$.
This leaves us to consider the density dependence;
a critical factor.

Knowledge of the density (or chemical-potential) dependence,
essential to transconductance, must be inferred from the
Fermi-liquid nature of the carriers.
In a degenerate channel the field acts bodily on every carrier
in the Fermi sea, but only the very few at its surface can scatter 
freely. Most are in lock-step. Thus the Fermi liquid as a whole
resists the tendency to emit phonons,
on par with its characteristic energy $E_{\rm F}$.

For $E_{\rm F} \gg eV$, Pauli blocking strongly
inhibits any and all increases in scattering rates.
We include the strong Fermi-liquid effects in Eq.\ (\ref{eq3})
by normalizing $eV$ to $E_{\rm av} \to E_{\rm F}$. Pauli
blocking is the main source of variation of
$\lambda_{\rm in}$ with $\mu$.
In the low-density limit (pinchoff), we have
$E_{\rm av} \to k_{\rm B}T$, so $\lambda_{\rm in}$
sheds its dependence on density.

Figure 3 compares our results from Eqs.\ (\ref{eq1})--({\ref{eq3})
directly with experimental data for different length channels.
The quality of fit to the measured $I$--$V$ curves
\cite{pop}
is good. The asymptotic tails are faithfully rendered.
In the shortest channels we see some departure
from measured data. We ascribe this to
(a) the proximity of the bulk leads,
which provide efficient heat sinking and thus qualitatively
change the hot-phonon behavior, and (b) nonuniform densities
in very short CNTs.

In signal detection, the transconductance is central.
We now address $g_m$, starting at Eq.\ (\ref{eq4})
and the in-channel response
${\widetilde g}_m \equiv e{\partial I/\partial \mu}$.
Since $I$ is a very strong function of
$\tau_{\rm in}(\mu, eV)$,
we see that ${\widetilde g}_m$ has a rich structure
wherein $\mu$ and $V$ enter more than once:

\begin{widetext}
\vspace{-7mm}
\begin{equation}
{\widetilde g}_m(\mu, eV)
= e{\partial I\over \partial \mu}
= eI {\partial\over \partial \mu}{\left[
  \ln{\left( {\tau(\mu, eV)\over \tau_{\rm el}(\mu)} \right)}  
+ \ln{\left( {k_{\rm F}(\mu)\over k_{\rm av}(\mu)} \right)}
+ \ln{\left( 1 -
\exp{\left(-{\hbar L k_{\rm av}(\mu)\over
             eV\sqrt{\tau(\mu, eV)\tau_{\rm in}(\mu, eV)}} \right)}
\right)}
\right]}.
\label{eq5}
\end{equation}
\end{widetext}

Next we map ${\widetilde g}_m$ to $g_m$.
Our bias gate is a cylinder
of radius $R = 20$nm
coaxial with the CNT of diameter $d = 2.4$nm.
The capacitive loop comprises the bias
source $V_g$, the tube, and the gap to the gate. Thus
$\delta V_g = -\delta \mu/e - 2(-e\delta n)\ln(d/2R)/\epsilon$
where $\epsilon$ is the dielectric constant and $-e\delta n$
is the induced change in electron charge density.

The bias loop's inverse total differential capacitance is
$C_g^{-1} = -\delta V_g/(eL\delta n)
= (2/\epsilon L)\ln(2R/d) + C_q^{-1}$ and
the term $C_q = e^2 L\partial n/\partial \mu$
is the ``quantum capacitance''
\cite{pulfrey,kimgm}.
Then
\vspace{-3mm}

\begin{equation}
g_m
= -{\delta I\over \delta V_g}
= e{\partial I\over \partial \mu}
{\left(  { {\delta \mu/(e^2 L\delta n)}\over
           {-\delta V_g/(eL\delta n}) } \right)}
= {\widetilde g}_m {C_g\over C_q}.
\label{eq6}
\end{equation}

\noindent
In metallic CNTs the large group velocity greatly
enhances $C_q$ so that  $|C_g/C_q - 1| \lesssim 10^{-3}$.
Hence $g_m = {\widetilde g}_m$.

\begin{figure}
  \centering
\vspace{10mm}
  \includegraphics[height=70mm]{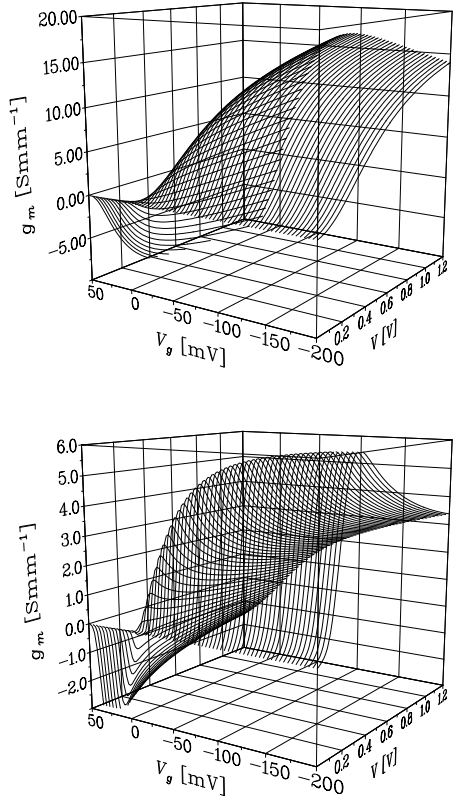}
\vspace{25mm}
  \caption{\label{f4}
Total transconductance (electronic plus hole) in a zero-gap
metallic nanotube, as a function of gate bias $V_g$
and source-drain voltage $V$, for the devices of Fig.\ 1.
Top: heat-sunk tube. Bottom: suspended tube.
In the latter, $g_m(V)$ tracks the current response,
but with even sharper overshoot. Over this range of $V$ and $V_g$,
the peak $g_m$ is 17Smm${}^{-1}$ for the heat-sunk device,
6Smm${}^{-1}$ for suspended.
}
\vspace{-5mm}
\end{figure}


A metallic CNT has zero band gap.
For good signal gain, however, we need
the operating region of maximal $g_m$;
that point lies well away from threshold
(where a semiconducting CNT would otherwise pinch off).
Moreover, since electron-hole recombination occurs on nanosecond
time scales, far longer than transit times $\sim$ 6ps
over 3$\mu$m, hole and electron currents simply add
and each is described by Eq.\ (\ref{eq4}).
Thus, in terms of $g_{m;e}$ and $g_{m;p}$, its electron and hole parts, 
$g_m = g_{m;e} + g_{m;p}$.

Figure 4 displays our predicted $g_m$ for the same
heat-sunk (upper panel) and free-standing (lower panel) CNTs
as in Fig.\ 1. Our kinetic model has been run with
the input data (refer to Fig.\ 2) extracted from our fits
to measurement
\cite{pop}.
Three features are notable.

{\em Maxima}. Each plot has a pronounced maximal ridge in
the electron-dominated region $\mu > 0$
(with a mirror valley, of opposite sign, in the hole-dominated
region $\mu < 0$).
For the heat-sunk device at all driving fields, $g_m$ peaks
somwhat above the band threshold.
By contrast, $g_m$ in the suspended tube has a maximal ridge
that is almost orthogonal to the former case.

{\em Overshoot}. As in Fig.\ 1, there is overshoot behavior in $g_m$
versus $V$ in the suspended sample at all $V_g$. The rapid rise
and fall of the transconductance with driving field is
more pronounced than for $I$ vs $V$.

{\em Suppression}. For the suspended tube, the transconductance
is suppressed threefold below $g_m$ in the heat-sunk tube.
This, and the overshoot, are of course intimately tied to the
drastic change in the dynamics of optical-phonon emission
between the two configurations.

We briefly discuss the cutoff frequency $f_{\rm T}$, above which
a transistor loses the capacity to follow a high-frequency gate
signal. It is given as $f_{\rm T} = g_mL/2\pi C_g$.
For the heat-sunk MSWNT of Fig.\ 4, $f_{\rm T}$ peaks at 140GHz
at $V_g = \pm 100$mV and $V = 1.2$V. For the suspended case,
the peak is 50GHz at $V_g = \pm 200$mV, $V = 0.35$V. 

In sum: we have used a microscopic Fermi-liquid kinetic analysis
to predict the transconductance of a metallic CNT, a property
vital to the design of high-gain nanotube transistors. Our theory is
strictly conserving, self-contained, and readily implemented.
To our knowledge, no detailed account of the critical
density dependence of $g_m$ has been available until now. 
Our predictions are directly testable in gated devices.
Systematic measurements of $g_m$ will allow new insights
into the dynamics of one-dimensional metallic channels.

Finally we note that suspended metallic nanotubes are extremely
unusual, in that both their high $g_m$ and their negative
differential conductance coexist in the same operating region.
Such a combination holds the clear possibility of a radically new
kind of active structure, with signal processing capabilities
previously unseen at nanoscopic scales.

We thank M. P. Das, H. Dai, P. McEuen, E. Pop,
and D. L. Pulfrey for valuable discussions.
This work was supported by the Australian Research Council.



\end{document}